\documentclass[12pt,aps,floatfix,nofootinbib]{revtex4}
\usepackage{amsmath}
\usepackage{latexsym}
\usepackage{float}
\usepackage{amssymb}
\usepackage{graphicx}
\usepackage{epsfig}

\newcommand{\theeq}{\theta_\mathrm{eq}}

\newcommand{\vslip}{v_\mathrm{slip}}

\begin{document}

\title{
Drop dynamics on hydrophobic and superhydrophobic surfaces
}

\author{B.\ M.\ Mognetti,$^{1}$  H.\ Kusumaatmaja,$^{2}$ and J.\ M.\ 
Yeomans,$^{1}$ \\
$^1$The Rudolf Peierls Centre for Theoretical Physics,
 1 Keble Road, Oxford, OX1 3NP, United Kingdom.\\
$^2$Department of Theory and Biosystems, Max Planck Institute of Colloids and 
Interfaces, 14424 Potsdam, Germany.
}

\begin{abstract}

We investigate the dynamics of micron-scale drops pushed across a
hydrophobic or superhydrophobic surface. The velocity profile 
 across the
drop varies from quadratic to linear with increasing height, 
indicating a crossover from a sliding to a rolling motion. We identify 
a mesoscopic slip capillary number which depends only on the motion 
of the contact 
line and the shape of the drop, and show that the
angular velocity of the rolling increases with increasing viscosity.
For drops on superhydrophobic surfaces we argue that a tank treading 
advance from post to post replaces the diffusive relaxation that allows 
the contact line to move on smooth surfaces. 
Hence drops move on superhydrophobic surfaces more quickly than on 
smooth geometries.

\end{abstract}

\maketitle

electronic mail: b.mognetti1@physics.ox.ac.uk

\section{Introduction}

The question of how liquid drops move across a solid surface has long caught 
the interest of academic and 
industrial communities alike, with applications ranging from microfluidic 
devices to fuel cells and inkjet 
printing. In many cases, efficient and effective control of the drop dynamics 
is highly desirable, and this 
relies upon understanding the internal fluid motion of the drop.

It has been reported in the literature 
\cite{Kim, EXP1-ROLL, EXP2-ROLL, 1, Richard-99} that a liquid drop 
may move in a  
variety of ways including sliding, rolling, tank treading and slipping. 
In some cases, a 
pearling instability may also be observed at the trailing edge of the drop
\cite{PFL-01}. 
 Highly viscous drops roll rather than slide \cite{Richard-99},
and on superhydrophobic surfaces drops appear to move very easily 
\cite{supH}. Using small particles as tracers, it 
has recently become possible  to access the velocity profile within a
 drop \cite{Richard-99,1,EXP1-ROLL,EXP2-ROLL}. However it is 
not yet always clear how the internal fluid motion is related 
to physical parameters such as fluid 
viscosity, the strength of the forcing, the equilibrium contact 
angle and hysteretic properties of the surface. Controlled experiments 
are typically possible only over a 
restricted range, and can be difficult for  small drops of size
 below the capillary length. Moreover, 
certain parameters, such as surface roughness, can be difficult to control 
and yet may have an important
role. Analytical calculations are possible, but they are typically 
limited to  small \cite{Durbin, Thiele} and high 
\cite{MP-99,HJR-04} contact angles. Computer simulations are 
therefore highly desirable 
to bridge the knowledge from experiments and analytical theories.

Several simulation methods have been developed to shed light on 
the dynamics of moving drops 
\cite{LB-droplet1,LB-droplet2,SM-08, flying, driven2, Spelt, driven3}.  
The  computational tool we use here is a mesoscopic diffuse interface model 
\cite{Swift,BriantBinary} solved using a lattice Boltzmann algorithm 
\cite{Yeomans,Succi}. This proves a useful approach due to its ability to 
handle interfacial dynamics and complex geometries easily.  For example, it 
has recently been used to study capillary filling \cite{Chirabbo-09},
hysteresis of drops on patterned surfaces \cite{LB-droplet1.5,Blow-08},
instabilities and detaching phenomena \cite{flying} and driven drops
in external flows \cite{driven2,driven3}. Diffuse interface models are 
mesoscopic approaches, which do not resolve the microscopic details of 
the contact line dynamics. As such, they provide a useful complement to 
molecular dynamics simulations.

We concentrate on modelling micron-scale drops moving on hydrophobic and superhydrophobic surfaces. We find that, for both hydrophobic and superhydrophobic substrates, the velocity profile across the height of the drop falls into two regimes. Near the substrate the velocity varies quadratically with height corresponding to a sliding motion. This can be characterized in terms of a mesoscopic slip capillary number which depends on the equilibrium contact angle and the mobility of the contact line. Further from the surface the velocity varies linearly with height indicating rolling. There is no dependence of the angular velocity on microscopic details of the contact line motion, such as the mobility, but the rolling becomes faster with increasing viscosity  for a given body force $f$.

We identify several major differences between the motion on hydrophobic surfaces, which are smooth, and similar surfaces which are patterned with posts such that they become superhydrophobic.  On the superhydrophobic substrates pinning of the contact line at the edges of the posts means that there is a threshold forcing below which the drop will not move.  However, once they do start to move, they do so much more easily and drops can reach much larger velocities before being detached from the surface in agreement with the experimental observations. We impose no-slip boundary conditions in the simulations, and on the smooth surfaces there is indeed no slip at the surface. However, the fluid velocity at the surface is non-zero and independent of the mobility for drops moving on superhydrophobic surfaces \cite{sh0,sh1,sh2} suggesting that the interface moves by tank treading between successive posts, rather than by diffusion-mediated motion of the interface.

The plan of the paper is as follows: In Sec.\ \ref{secM} the numerical 
model is introduced,
and in Sec.\ \ref{secSP} the simulation set-up and the computational 
parameters  are given. 
Considering first  drops on smooth surfaces, in Sec.\ \ref{SecCrossover} 
we describe the 
crossover from a Poiseuille-like flow to a pure rolling regime 
with increasing height and point
out the relevance of the equilibrium contact angle. In Secs.\ \ref{SecVslip} 
and \ref{SecOmega} we investigate the dependence of the Poiseuille and 
rolling regimes on the viscosity and surface tension of the fluid  
and on the mobility of the contact line. In Sec.\ \ref{SecPatt} 
we compare the dynamics of drops on superhydrophobic surfaces. Finally, in 
Sec.\ \ref{SecConclusions}, we summarise and discuss our conclusions.

\section{The model }\label{secM}

We use a binary fluid model, 
with components $A$ and $B$, described by a Landau  free energy 
 \cite{Swift,BriantBinary}
\begin{eqnarray}
\Psi = \int_V (\psi_b + \frac{\kappa}{2} (\partial_{\alpha}\phi)^2) dV + \int_S \psi_s \, dS
\label{freeen}
\end{eqnarray}
where the bulk free energy density $\psi_b$ is taken to have the form
\begin{eqnarray}
\psi_b = \frac{c^2}{3} n \ln n + A \left( -\tfrac{1}{2} \phi^2 + \tfrac{1}{4} \phi^4 \right).
\label{bulk}
\end{eqnarray}
The functional (\ref{freeen}) is discretised on a cubic lattice 
with lattice spacing $\Delta{}x$, and $\Delta{}t$ is the 
simulation time step.
$n$ is the total fluid density $n=n_A+n_B$,   $\phi$ is the order parameter $\phi=n_A-n_B$,
and  $c=\Delta{}x/\Delta{}t$. This choice of $\psi_b$ gives binary phase 
separation into two phases with $\phi = \pm 1$. 

The gradient term in Eq. \eqref{freeen} represents an energy contribution from variations in $\phi$ and is related to the surface tension between the two phases by $\gamma = \sqrt{8\kappa A/9}$  and to the interface width through $\xi = \sqrt{\kappa/A}$  \cite{BriantBinary}.

The second integral in Eq. \eqref{freeen} is taken over the system's 
solid surface 
and is responsible for the wetting properties. 
The surface energy density is chosen to be $\psi_s = -h \phi_s$ \cite{Cahn}, where $\phi_s$ is the value of the order parameter at the surface. Minimisation of the free energy shows that in equilibrium the gradient in $\phi$ at the solid boundary is \cite{BriantBinary}
\begin{equation}
\kappa \partial_\perp \phi =  -\frac{d\psi_s}{d\phi_s} = - h \, .
\end{equation}
The value of the phenomenological parameter $h$ is related to the equilibrium contact angle $\theeq$ ($\theeq$ is taken with respect to the $\phi=1$ component) via \cite{BriantBinary}:
\begin{eqnarray}
h  = \sqrt{2\kappa A} \,\, \text{sign} \left(\frac{\pi}{2} \!- \! \theeq \right)  \sqrt{\cos\left(\frac{\alpha}{3} \right) \left\{ 1 \!- \! \cos\left(\frac{\alpha}{3}\right)\right\}} \, ,
\label{theory}
\end{eqnarray}
where $\alpha = \cos^{-1} \left(\sin^2 \theeq \right)$ and the function sign returns the sign of its argument. 

The hydrodynamic equations for the binary fluid are 
\begin{eqnarray}
&\partial_{t}n+\partial_{\alpha}(nv_{\alpha})=0 \, , \label{ch1eq5}\\
&\partial_{t}(nv_{\alpha})+\partial_{\beta}(nv_{\alpha}v_{\beta}) = 
- \partial_{\beta}P_{\alpha\beta}+ \partial_{\beta}[\eta (\partial_{\beta}v_{\alpha} 
+ \partial_{\alpha}v_{\beta} + \delta_{\alpha\beta} \partial_{\gamma} v_{\gamma}) ] 
+  f_{\alpha} \, , \label{ch1eq6} \\
&\partial_t{\phi} + \partial_\alpha \left( \phi v_\alpha \right) = M \nabla^2 \mu. \label{nsfinal2}
\end{eqnarray}
Eq.\  \eqref{ch1eq5} is the continuity equation, Eq.\ \eqref{ch1eq6}  the 
Navier-Stokes equation and Eq.\ \eqref{nsfinal2}  the convection diffusion 
equation.  $M$ is a mobility coefficient. 
In Eq.\ (\ref{ch1eq6}) we have introduced the dynamic viscosity $\eta$, 
and a bulk force for unit volume $f_\alpha$.
The equilibrium properties of the fluid appear in the equations of motion through the pressure tensor and the chemical potential. Both can be obtained in the usual way from the free energy and are given by \cite{BriantBinary}
\begin{eqnarray}
&\mu  = A\left(- \phi + \phi^3 \right)- \kappa \nabla^2 \phi, \label{chempot} \\
&P_{\alpha \beta} =  \left(p_b - \frac{\kappa}{2}(\partial_\gamma \phi)^2 - \kappa \phi \partial_{\gamma\gamma} \phi \right) \delta_{\alpha \beta} + \kappa (\partial_{\alpha} \phi)  (\partial_{\beta} \phi) \, , \label{pressten} \\
&p_b =  \tfrac{c^2}{3}n  +  A \left( -\tfrac{1}{2} \phi^2 + \tfrac{3}{4} \phi^4 \right). \nonumber
\end{eqnarray}

To solve Eqs. (\ref{ch1eq5}), (\ref{ch1eq6}) and (\ref{nsfinal2}), we use the 
free energy lattice Boltzmann scheme recently presented by Pooley 
{\em et.\ al.\ } 
\cite{PooleyMRT}. The main advantage of this method is that the spurious 
velocities at the contact line are strongly reduced. This is particularly 
important when the two fluid components have different viscosities.

A word of explanation is in order as to why  we are using a binary fluid 
approach to describe a liquid-gas system.
Recently several authors \cite{Pooley-09,Chirabbo-09,Diotallevi} have 
investigated capillary filling
using liquid-gas and binary lattice Boltzmann algorithms. They found that 
the binary model is able to reproduce the experimental
results, but that the liquid-gas model produces filling that
is faster than expected. This is due to excessive condensation
of the gas phase at the interface. The effect is suppressed only when 
the liquid
density is much larger than the gas density, a regime difficult to access 
numerically. Therefore we use a binary model. However, our results are equally 
applicable to a physical system where a liquid displaces a gas  if the 
important physical parameters are the
viscosities,  rather than the densities of the fluid components.
For convenience we will use liquid/gas terminology in the rest of the
paper.

The way in which the contact line moves does depend on the choice of model. 
Here it is determined by the mobility $M$ which determines how 
the diffusion across the interface is driven by variations of the chemical
potential $\mu$. Given that the model does not include microscopic 
details we can say nothing about the mechanisms operating at the contact line 
in a physical system, and we regard $M$ as a phenomenological parameter which 
controls the rate of slip.

\section{Geometry and simulation parameters}\label{secSP}

We consider drops of size smaller than the capillary length, so that we can ignore gravity, and sufficiently large that thermal fluctuations can be neglected. For a water drop this corresponds to a range of length scales between $\sim 10^{-6}$m and $ 10^{-3}$m.  On these length scales the motion of the drop depends on its viscosity $\eta$ and surface tension $\gamma$ and a useful dimensionless measure of drop velocity $v$, allowing direct comparison to experiments, is the capillary number $Ca=v\eta/\gamma$.
In the simulations we use three different values for the liquid viscosities ($\eta=0.5$, 0.833, 1.1667) and two for the surface tension
($\gamma=0.0267$, 0.0533). The gas viscosity $\eta_{gas}$ is always chosen to keep the ratio $\eta / \eta_{gas}=25$ and a
 linear interpolation  between $\eta$ and $\eta_\mathrm{gas}$ is used to define the viscosity variation through the interface.

A body force $f$ is applied to the liquid phase. (For example, for a drop moving down an inclined plane with tilt angle $\alpha$,
$f=n\cdot g\cdot \sin\alpha$ where $g$ is the acceleration due to gravity.) 
$f$ is taken to vary with $(1+\phi({\bf r}))/2$ which provides a linear interpolation from
$f$ to 0 when moving from the liquid to the gas. Other interpolating functions (e.g.\ $\tanh$) were tested, 
without significant differences.
Driving by the body force is normally reported in terms of a dimensionless
Bond number $Bo=f \cdot V/(\gamma \cdot L_y)$, where $V$ is the volume of the 
drop, $L_y$ is the length of the cylinder which is equal to the length of the 
simulation
box in the $y$-direction, and $f$ is the body force defined in 
Eq.\ (\ref{ch1eq6}). 
We choose values of $f$ between $0.05\cdot 10^{-6}$ and $1.6\cdot 10^{-6}$ 
and $V=11 \cdot 10^3 \cdot L_y$.

Other parameters important in determining the drop motion are the mobility $M$, where we use values from $M=0.125$ to $M=2$,
and the equilibrium contact angle $\theta_{eq}$, considered below.

\begin{center}
\begin{figure}
\vspace{0.5cm}
\includegraphics[angle=0,scale=0.5]{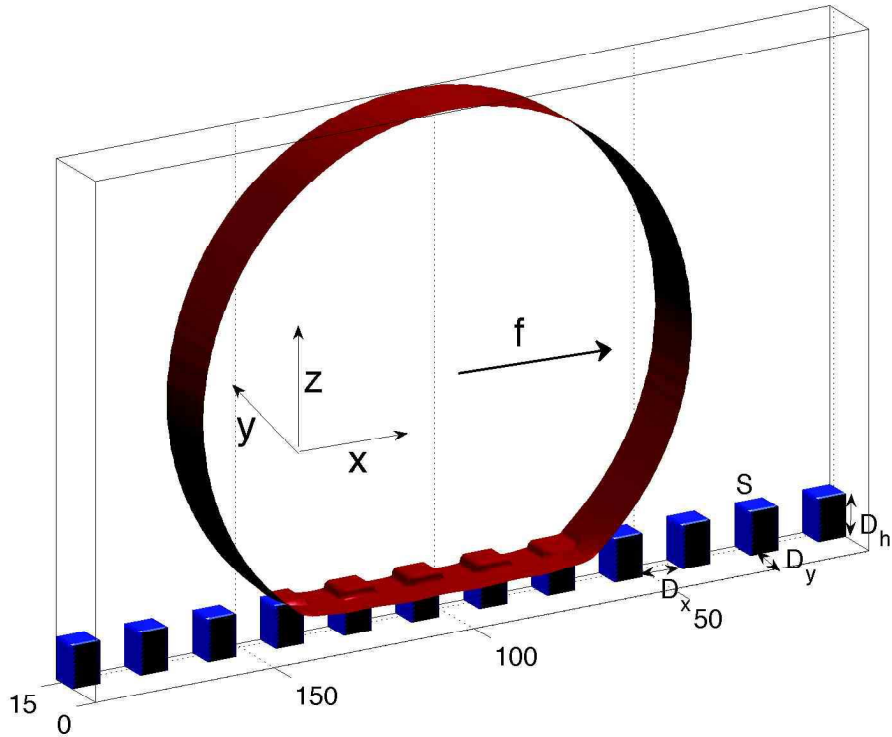}
\vspace{0.5cm}
\caption{Simulation geometry for the patterned substrate. A cylinder
of fluid is prepared in the suspended state and a force 
per unit volume  f  is 
applied along the $x$ direction. }\label{fig1}
\end{figure}
\end{center}
We model cylindrical drops placed on a smooth surface or
on a surface decorated by regular posts as shown in  Fig.\ \ref{fig1}. For the patterned substrates
the use of cylindrical drops is a compromise between two dimensional simulations, which do not 
well describe the pinning of the contact line on the posts (as they essentially replace the posts by ridges), and full three dimensional simulations which are prohibitively time consuming.

For the smooth surfaces we mainly use a hydrophobic equilibrium contact angle 
$\theeq = 145^\circ$, but results at other contact angles are presented for comparison.
For the patterned surface  the posts were taken to have a square cross section 
of side $S=8\cdot \Delta x$, height 
$D_h=10\cdot \Delta x$, and to be equispaced 
by $D_x=D_y=9$ lattice units in both plane directions 
(see Fig.\ \ref{fig1}). A hydrophobic surface patterned by posts becomes superhydrophobic. 
There is a large increase in the contact angle $\theta_C$ of a drop resting on the posts 
given by the Cassie formula $\cos \theta_C=\phi \cos \theeq+\phi-1$ \cite{Cassie} where $\phi$ is the area
fraction of the flat surface  covered by posts. For the geometry we consider
$\phi=S^2/(S+D_x)^2\approx 0.22$. Hence
using  $\theeq=100^\circ$ gives a Cassie angle $\theta_C\approx 145^\circ$
equal to the $\theeq$ used for the flat surface geometry.

As shown in Fig.\ \ref{fig1}, 
the substrate is placed in the $xy$ plane and 
the sliding motion is along the $x$-direction.
Typically we used a simulation box of size
$L_x\times L_y \times L_z = 180 \times 1 \times 180 $
(lattice units) for the smooth geometry, and $L_x\times L_y \times L_z 
= 204 \times 17 \times 180 $ (lattice units) when posts were present.
At small contact angles, for the flat plane, $L_x$ was sometimes increased 
to $L_x=360$ in order to minimize the interaction between the drop and 
its images created by the periodic boundaries conditions.
Cylindrical drops were initialized with a radius $R = 60$ centered $50$ 
lattice units 
above the plane (or the top of the posts in the suspended states)
and allowed to relax to equilibrium before the force was applied.

\section{Velocity profile inside a drop}\label{SecCrossover}

\begin{center}
\begin{figure}
\includegraphics[angle=0,scale=0.95]{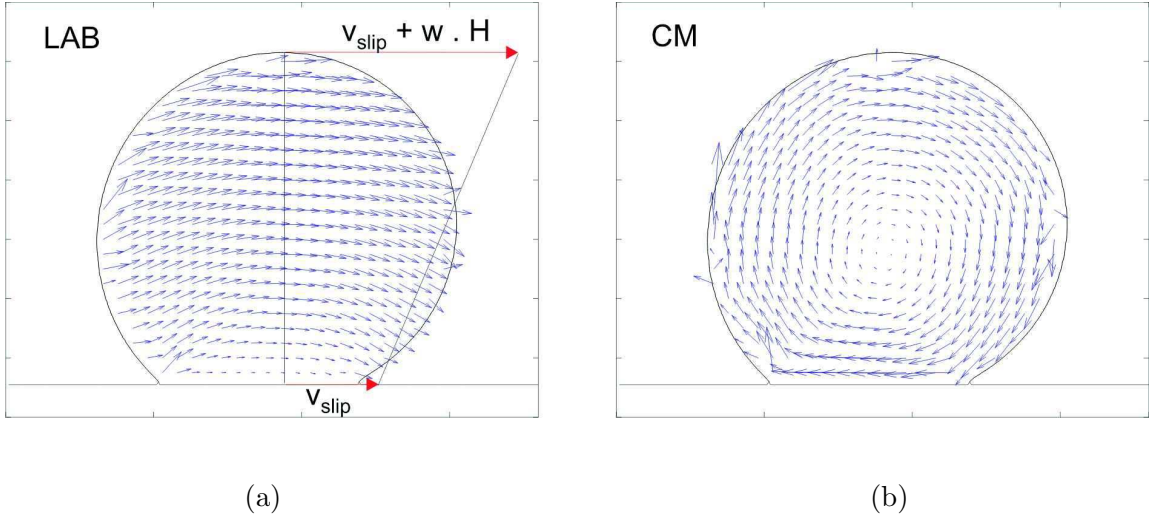}\\
\vspace{0.4cm}
(a) \qquad \qquad \qquad \qquad\qquad \qquad \qquad \qquad\qquad \qquad (b)\\
\vspace{0.4cm}
\caption{(Color online) Velocity profile for a drop with contact
angle $\theeq=145^\circ$
(a) in the laboratory reference frame and (b) in the drop's center of
mass reference frame. For clarity the velocities have been rescaled and 
we have not reported the flow of the surrounding gas. In panel (a) we 
illustrate the phenomenological 
 description of the motion of a drop as superposition of
a slip velocity $\vslip$ plus a rotational motion of angular velocity $w$.
}\label{FigVInside}
\end{figure}
\end{center}

Understanding the dynamics of a moving drop is more difficult 
than for rigid objects because of the possibility of the fluid shearing.
This means that, in general,  the drop motion will include both sliding 
and rolling components. Fig.\ \ref{FigVInside} reports a typical steady velocity 
profile found using the diffuse interface model Eq.\ (\ref{freeen}) for an 
equilibrium contact angle $\theeq=145^\circ$  in the laboratory and drop frames
 of reference.  In the latter
 (Fig.\ \ref{FigVInside}b) the presence of rolling is evident. 

\begin{center}
\begin{figure}[h]
\vspace{1.5cm}
\includegraphics[angle=0,scale=0.5]{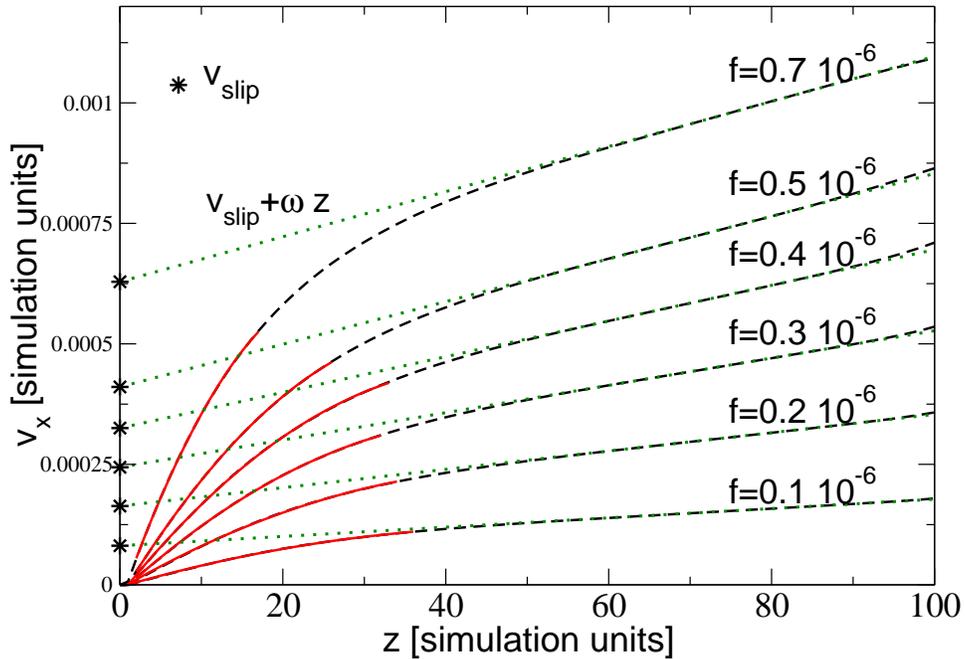}
\caption{(Color online) Steady velocity in the $x$ direction ($v_x$) 
as a function of the height $z$. At large $z$ a linear, rolling, regime
is present (dashed) which crossovers a quadratic profile for small $z$ 
(full lines).
$\vslip$ is  defined as the extrapolation of the linear regime to 
$z=0$. Data were obtained using $\eta=0.833$, $\gamma=0.02667$, 
$\theeq=145^\circ$ and $M=0.25$.}\label{FigCrossover}
\end{figure}
\end{center}

In Fig.\  \ref{FigCrossover} we report the $x$-component of the drop velocity $v_x$ as a function of height 
$z$ (computed  along the line denoted $H$ in Fig.\ \ref{FigVInside}) for different applied forces $f$. There 
is a clear separation into two regimes. For larger $z$ the velocity scales linearly with height, 
corresponding to rolling dynamics. For smaller $z$ the behaviour is close to quadratic and is well fit by a 
Poiseuille-like \cite{Landau} flow
\begin{eqnarray}
v_x(z) = v_0 + {f \tilde H \over  \eta}z - {1\over 2}{f\over \eta} z^2 \, \, , 
\label{EqVQuad}
\end{eqnarray}
where $\tilde H$ would correspond to the centre of the channel in the usual Poiseuille geometry
and $v_0$ allows for local slip at the liquid-solid interface. 
(For forces $f<0.5 \cdot 10^{-6}$, where the drops
are not too deformed, the quadratic term of the fit is in agreement with
the $z^2$ term of Eq.\ (\ref{EqVQuad}) to within 10\% .)
In our simulations we impose non-slip boundary conditions and therefore
 $v_0=0$ for smooth substrates by construction.  We 
will see later that this is no longer the case for superhydrophobic surfaces. 

Experimental results  \cite{1,EXP1-ROLL,EXP2-ROLL} are often fitted by
simplifying the drop dynamics as a superposition of a
rotation (with angular velocity $w$) and a constant sliding velocity 
$\vslip$. In this scheme the velocity profile in the $x$ direction
is a linear interpolation between $\vslip$ and the velocity at the top
of the drop (see the arrows in Fig.\ \ref{FigVInside}a)
\begin{eqnarray}
v_x &=& \vslip+ w\cdot z \, \, .
\label{EqVLinear}
\end{eqnarray}
Guided by the experimental approach, we identify the angular velocity $w$ as the slope of the linear  
 part of the velocity profiles. Extrapolating the linear fit to $z=0$ (see the dotted curves in Fig.\ 
\ref{FigCrossover}) we define a slip velocity $\vslip$ which characterizes the motion of the layers 
of fluid near the solid wall. 
(It is important to distinguish $\vslip$,   and the microscopic slip velocity $v_0$ which is the velocity at $z=0$.)  We will describe how $\vslip$ and how the rolling region of the velocity profile, parameterized by $\omega$, depend on parameters such as viscosity, surface tension, mobility and the shape of the drop in Sections \ref{SecVslip} and \ref{SecOmega} respectively.

\begin{center}
\begin{figure}[h]
\vspace{1.cm}
\includegraphics[angle=0,scale=0.5]{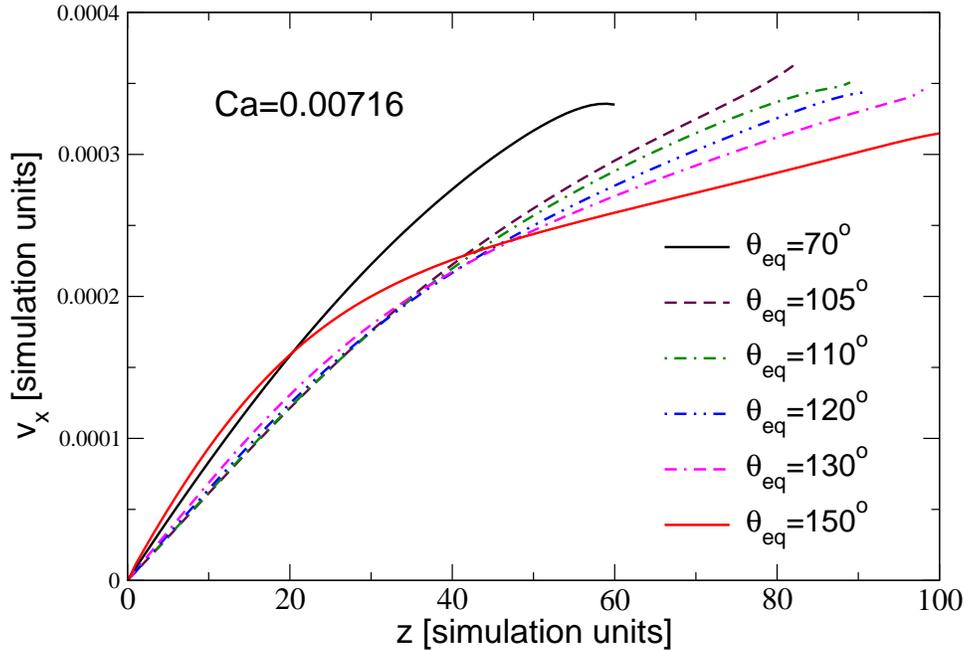}
\vspace{1.cm}
\caption{ (Color online) Variation of the velocity $v_x$ with 
height $z$  at different 
equilibrium contact angles but the same center of mass capillary 
number $Ca=v_\mathrm{cm}\eta/\gamma$
(where $v_\mathrm{cm}$ is the centre of mass 
velocity). Only for high equilibrium 
contact angles ($\theeq>120^\circ$) is a linear region clearly present.
The simulation parameters are the same as in Fig.\ \ref{FigCrossover}.
}\label{FigCrossThe}
\end{figure}
\end{center}

However first we present results showing that the equilibrium contact angle 
$\theeq$ strongly affects the 
crossover from the quadratic to the linear regime. In Fig.\ 
\ref{FigCrossThe} we report $v_x$ for different equilibrium contact angles
$70^\circ\leq \theeq \leq 150^\circ$ at the same centre of mass capillary 
number $Ca=v_\mathrm{cm}\eta/\gamma$, where $v_\mathrm{cm}$ is the centre of 
mass velocity. For $\theeq \lessapprox 120^\circ$, a
 linear part of the plot cannot be distinguished; 
only for $\theeq\gtrapprox 130^\circ$ are the two regions clearly 
visible. 
The results are consistent with lubrication theory, 
valid for small contact angles, which predicts a quadratic
profile \cite{Lub}, and scaling calculations 
that predict rolling at high contact angle \cite{MP-99,HJR-04}.
The applied force needed to produce the given capillary 
number in Fig. \ref{FigCrossThe}
increases with decreasing contact angle: for example the ratios 
are $1 : 1.7 :\hspace{0.00cm} 2.7$ for
$\theeq=70^\circ$, $105^\circ$ and $150^\circ$. The dissipation is larger for drops of smaller contact angle because the region in which the velocity varies quadratically with height extends further from the substrate.

\section{Slip velocity,  $\vslip$}\label{SecVslip}

\begin{center}
\begin{figure}
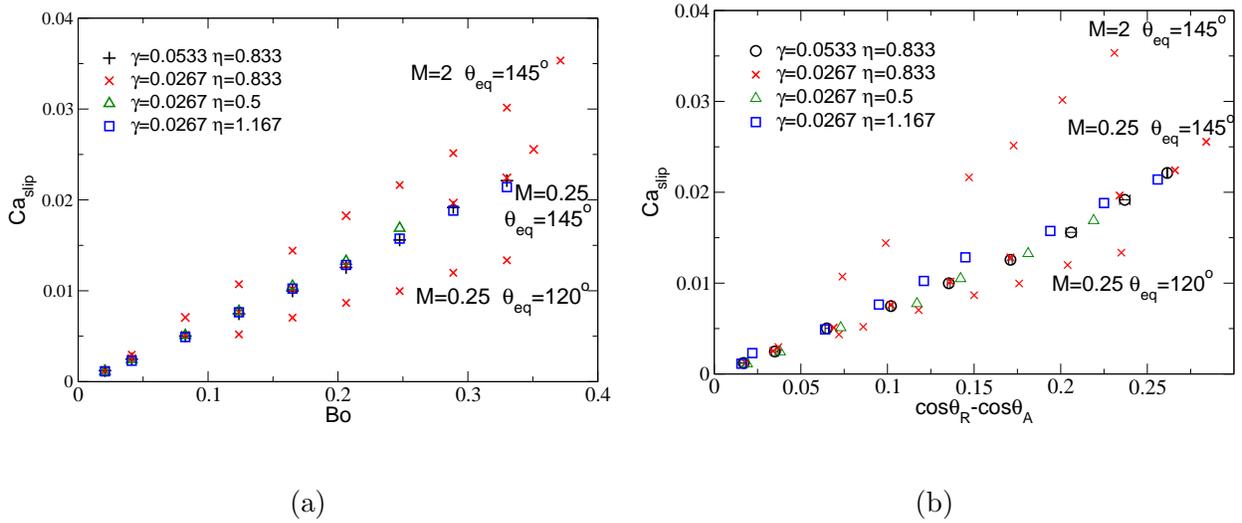

\vspace{1.cm}
 \includegraphics[angle=0,scale=0.32]{Fig.5a.eps}\, \,
 \includegraphics[angle=0,scale=0.32]{Fig.5b.eps} \\
\vspace{0.4cm}
(a) \qquad \qquad \qquad \qquad \qquad \qquad \qquad \qquad 
\qquad \qquad  (b) \\
\vspace{0.4cm}
\caption{(Color online)  Dimensionless  slip velocity $Ca_\mathrm{slip}$ 
versus (a) the Bond number $Bo$  and (b) $\cos\theta_R-\cos\theta_A$ where
$\theta_R$ and $\theta_A$ are the receding and advancing contact angles.
The curves depend on the mobility $M$ and equilibrium contact angle $\theeq$
but not, to with the precision of our data, the surface tension $\gamma$
or the viscosity $\eta$.
}\label{FigVslip}
\end{figure}
\end{center}

We now discuss the dependence of the mesoscopic slip velocity  on the parameters of the model. Fig.\ \ref{FigVslip} shows a dimensionless measure of $\vslip$, $Ca_\mathrm{slip} = \vslip \cdot \eta/ \gamma$, plotted against a dimensionless measure of the forcing, the Bond number, $Bo= f \cdot V / (\gamma \cdot L_y)$. There is a data collapse onto curves which depend only on the mobility $M$ and on the equilibrium contact angle $\theeq$, 
and $Ca_\mathrm{slip}=f(\theeq,M)\cdot Bo$ holds well for $Bo$ sufficiently small that the drop is not 
significantly deformed. As the mobility $M$ increases $\vslip$ increases for a given Bond number as 
expected, as it becomes easier for the contact line to relax. For smaller equilibrium contact angles $\vslip$ 
decreases for a given Bond number as the dissipation of the drops increases.

Another property of the moving drop that is related to $\vslip$ is the
degree of deformation caused by the forcing. In Fig.\ \ref{FigVslip}(b) we present data that suggest that, 
for small forcing, $v_\mathrm{slip}$ is linear in the uncompensated Young stress
$\Delta\equiv \gamma (\cos \theta_R - \cos \theta_A)$ where 
$\theta_R$ and $\theta_A$ are the receding and advancing contact angles. Again the slopes of the curves 
depend on the mobility and the equilibrium contact angles, but not on the viscosity or surface tension. We 
caution that the data here is noisy, as the deviations from the equilibrium contact angles are small and the algorithm, for high viscosities,  does not precisely reproduce $\theeq$ \cite{PooleyMRT}.

It is plausible and consistent with our results that the mobility coefficient $M$
affects $Ca_\mathrm{slip}$ only through the deformation of the droplets,
in such a way that a more detailed scaling relation for $Ca_\mathrm{slip}$  could
be written $Ca_\mathrm{slip}=\alpha(\theeq)\cdot Bo+\beta(\theeq,M) \cdot \Delta(Bo)$.
Qian {\it et al.} \cite{QWS-03-04,QWS-06} have
also highlighted the importance of
$\Delta$ in controlling interface motion. However the relation between our work and theirs is not yet clear as they consider a 
local microscopic  slip near the interface, whereas our parameter $\vslip$ follows from extrapolating from the 
bulk velocity profile. 
Work is in progress to investigate this further.

\section{Rolling}\label{SecOmega} 

\begin{center}
\begin{figure}
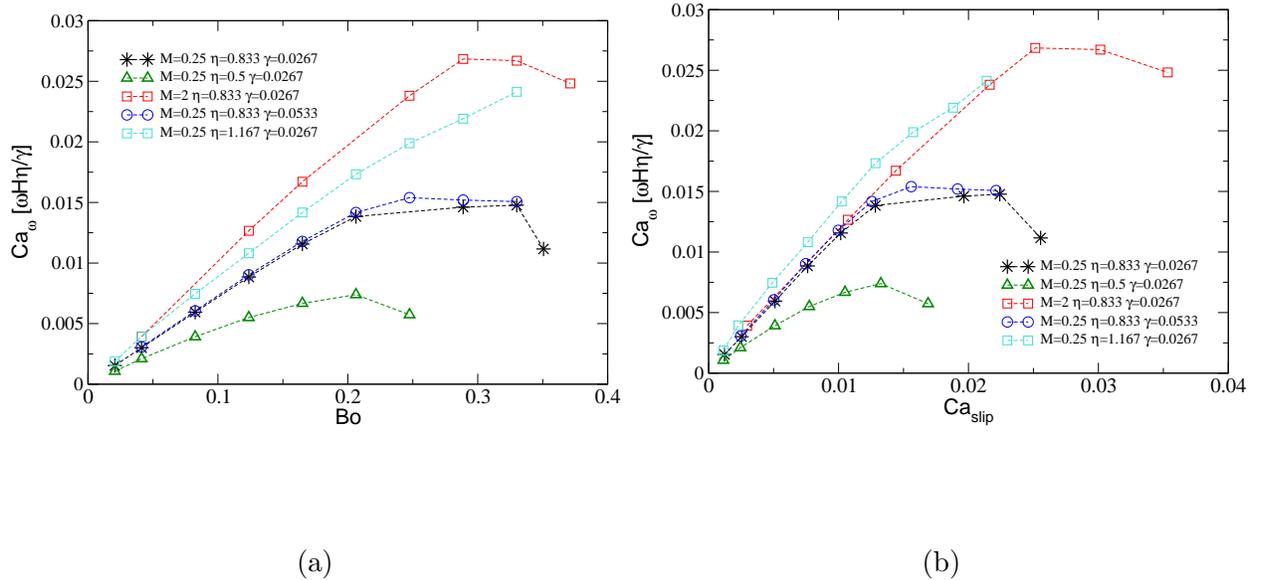

\vspace{1.cm}
\includegraphics[angle=0,scale=0.32]{Fig.6a.eps}
\includegraphics[angle=0,scale=0.32]{Fig.6b.eps} 
\vspace{0.4cm}\\
(a) \qquad \qquad \qquad \qquad \qquad \qquad \qquad \qquad 
\qquad \qquad  (b) \\
\vspace{0.4cm}
\caption{(Color online) Dimensionless angular velocity 
$Ca_w=w\eta H/\gamma$  against (a) Bond number $Bo$ 
(b) dimensionless slip velocity $Ca_\mathrm{slip}=\eta\vslip/\gamma$.
Curves for  different surface tensions 
(but equal viscosity)  collapse, while  $Ca_w$ 
increases with $\eta$.  In (b) the curves are also independent of mobility.
}\label{FigW}
\end{figure}
\end{center}

The rolling motion displayed away from the surface by drops of 
higher equilibrium contact angle has a very different dependence on 
the model parameters. The rolling 
contribution to the velocity profile of the drop is $w \cdot z$ in Eq.\
(\ref{EqVLinear}), corresponding to the linear part 
of the velocity profiles in Fig.\ \ref{FigCrossover}. We estimate a typical rolling velocity as $w\cdot H$, where $H$ is a 
length comparable to the height of the droplet.
A capillary number associated with the rolling motion can then be defined as 
$Ca_w=w\cdot H \cdot \eta/\gamma$. For the graphs in Fig.\ \ref{FigW}
we have consistently used $H=100$ lattice units.

In Fig.\ \ref{FigW} we plot $Ca_w$ as a function of the Bond number for different mobilities, viscosities 
and surface tensions. Fig.\ \ref{FigW}a shows that different viscosities 
do not collapse on a single master curve but rather that
$Ca_w$ increases with $\eta$. This means that the sliding and the rolling motion 
of the drops are affected in a different way by the viscosity. In particular the rolling 
component is favoured for high viscosity. This is not the case
for the surface tension $\gamma$: data for drops with different $\gamma$
(but equal $\eta$) collapse onto the same curves to within 
the precision of the data.

In Fig.\ \ref{FigW} $Ca_w$ is also larger for sets of data with
high mobility. This is not unexpected because $Ca_\mathrm{slip}$
is also $M$ dependent (Fig.\ \ref{FigVslip}a).
To understand if the mobility affects the rolling part of the motion,
we need to compare $Ca_w$ at fixed $Ca_\mathrm{slip}$. This is done
in Fig.\ \ref{FigW}(b) where it is apparent that different 
mobilities collapse onto
the same curves showing that $M$ affects the rolling part
of the motion only through $\vslip$.

For higher forcing the linear dependence between the capillary number 
and the Bond number is lost.
$\omega$ decreases,   $\vslip >> w \cdot H$,  and sliding dominates the motion.
This corresponds to a regime where the drops are highly deformed and a further 
increase in Bond number causes them to detach from the substrate \cite{flying}.
Similar behaviour has been observed
in experiments \cite{EXP2-ROLL} for drops on hydrophobic surfaces.

\section{Drops on patterned surfaces}\label{SecPatt}

In this section we investigate the dynamics of a drop
on a surface patterned with  posts (see Sec.\ \ref{secSP} 
and Fig.\ \ref{fig1} for  the geometric details). 
In particular we are interested in understanding analogies to, and differences from, 
motion across a smooth surface. Drops which lie on top of the posts in the suspended or Cassie state  
exhibit an apparent contact angle $\theta_C$ 
given by Cassie's relation $\cos \theta_\mathrm{C}=
\phi \cos \theeq + \phi-1$ \cite{Cassie}, where $\theeq$ is the equilibrium 
contact angle for the smooth plane. 
We consider cylindrical drops on surfaces patterned 
by obstacles which occupy an area fraction $\phi=0.22$ of the flat surface,
 and a value
$\theeq=100^\circ$, which gives $\theta_C\approx 145^\circ$, allowing 
a direct comparison with the results for a smooth surface in Secs.\ \ref{SecVslip} and \ref{SecOmega}.

In presence of posts  the drop remains stationary  for small 
Bond numbers  \cite{deGennes,HY-07,Reyssat-09}. 
This occurs because the interface 
is pinned at the edges of the obstacles \cite{Gibbs}. For the geometry we 
consider here the depinning threshold $Bo_T \sim 0.15$. 
For higher Bond number we observe oscillations in the velocity
which reflect the position of the drop relative to the posts and
become smaller further from the pinning threshold $Bo_T$.
The steady velocities were therefore computed by averaging over 
 a time interval much longer than the time required for the drop to
cross a post.

\begin{center}
\begin{figure}
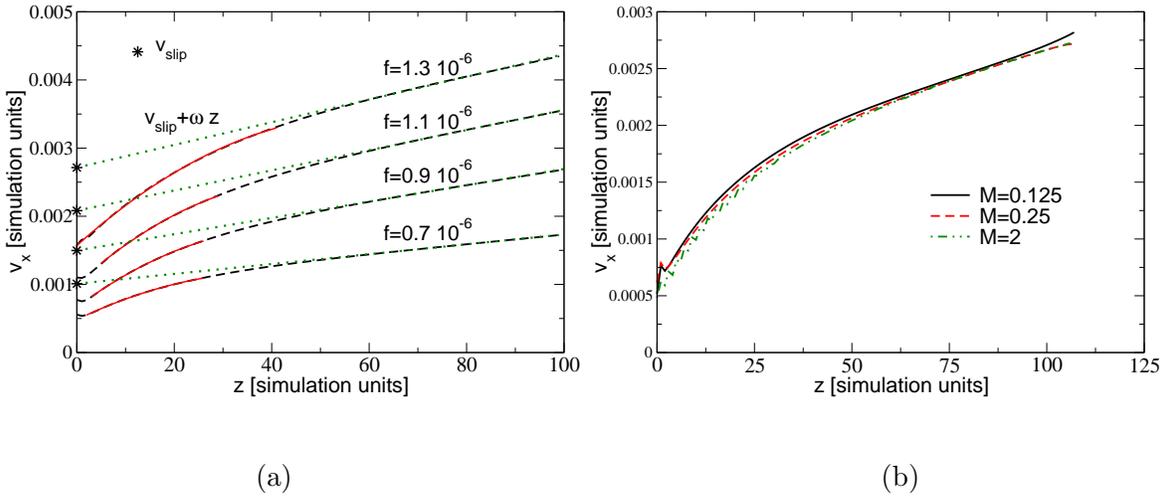

\vspace{1.cm}
\includegraphics[angle=0,scale=0.3]{Fig.7a.eps} 
\includegraphics[angle=0,scale=0.3]{Fig.7b.eps} 
\vspace{0.4cm}\\
(a) \qquad \qquad \qquad \qquad \qquad \qquad \qquad \qquad \qquad \qquad (b)\\
\vspace{0.4cm}
\caption{(Color online)  (a) The same as Fig.\ \ref{FigCrossover} but 
for superhydrophobic surfaces and for $0.29<Bo<0.54$. 
The plot shows the velocity in the $x$ direction ($v_x$) 
as a function of the height $z$  (measured from the top of
the posts). At large $z$ a linear, rolling, regime
is present (dashed) with crossover to a quadratic profile for small $z$ 
(full lines). (b) $v_x$ for Bond number $Bo=0.371$ ($f=0.9\cdot 10^{-6}$) 
and three
different mobilities. 
 Data were obtained using $\eta=0.833$ and $\gamma=0.02667$.
}\label{figVX}
\end{figure}
\end{center}
In Fig.\ \ref{figVX}(a) we plot the velocity profile $v_x$ as a function of the distance from the 
top of the posts for Bond numbers $0.29<Bo<0.54$. This plot should be compared to Fig.\ \ref{FigCrossover}
for the smooth surface. Just as for the smooth surface, the velocity profile 
exhibits a crossover from a quadratic to a linear regime. The
main difference is the presence of a microscopic slip velocity at  $z=0$, defined as  $v_0$
in Eq.\ (\ref{EqVQuad}), for the superhydrophobic substrate
\cite{sh1,sh2}. These results show that 
a crossover from 
a quadratic to a linear dependence of velocity on $z$  is not dependent on having zero  $v_0$
but is also present if some slip occurs at the solid--liquid interface. We remark that it is difficult to obtain a precise value for 
$v_0$ from the simulations because of the presence of spurious velocities at the surface.

An important difference between the smooth and the patterned geometry is
illustrated in Fig.\  \ref{figVX}(b) where we plot $v_x$ against $z$ for 
$Bo=0.371$ and different values of the mobility $M$.
The velocity is unaffected by the mobility, in stark contrast to the smooth surface where $Ca_\mathrm{slip}$ depends strongly 
on $M$ (compare Figs.\ \ref{FigVslip}). 
 This provides evidence that the contact line is moving onto successive posts using a tank-treading mechanism, rather than through relaxing the interface distortion by diffusion.

\begin{center}
\begin{figure}
\includegraphics[angle=0,scale=0.5]{Fig.8a.eps} 
\\
\vspace{0.5cm}
(a)\\
\vspace{1.4cm}
\includegraphics[angle=0,scale=1.00]{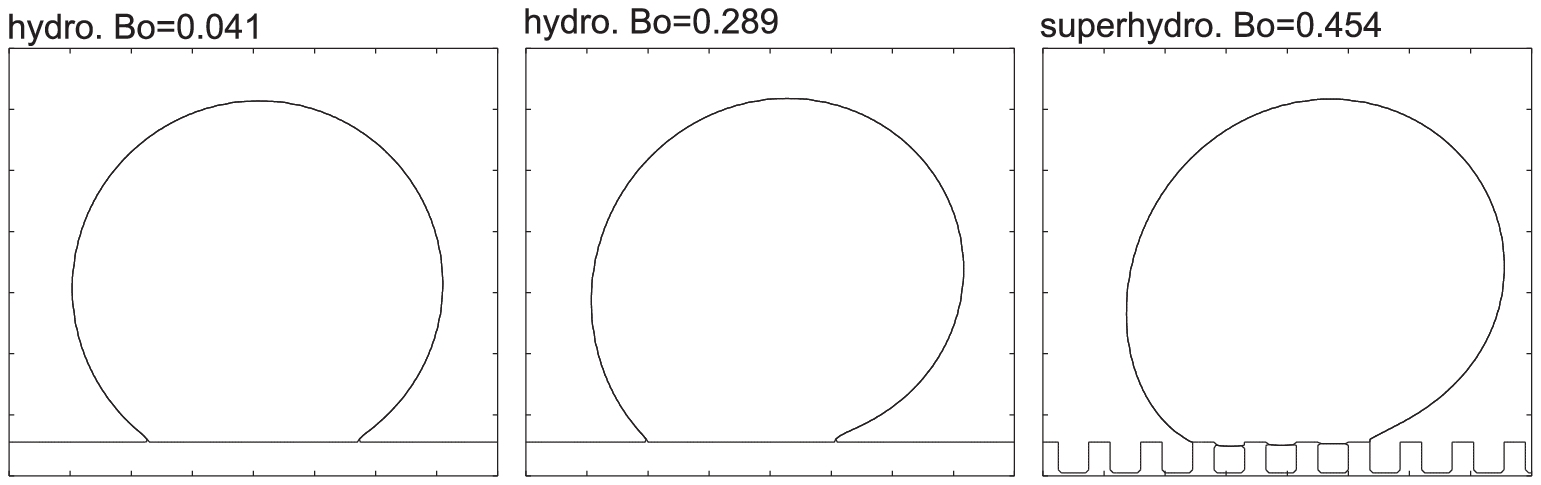} \\
\vspace{0.5cm}
(b)
\caption{(color online) (a) Drop center of mass capillary number 
$Ca_\mathrm{cm}$versus 
Bond number $Bo$ for drops on hydrophobic and superhydrophobic surfaces. 
 The equilibrium
contact angle of the smooth surface  is $\theeq=145^\circ$, while for the
patterned plane it is $\theeq=100^\circ$ giving a Cassie angle 
$\theta_\mathrm{c}\approx145^\circ$. (b) Steady 
state shapes of the drops at different $Bo$. }\label{figvcm}
\end{figure}
\end{center}

Fig.\ \ref{figvcm} is a plot comparing the centre of mass capillary number as a function of Bond number for drops on hydrophobic and superhydrophobic surfaces. Full lines refer to the smooth surface, while the broken line
is for the patterned geometry. As before, the equilibrium
contact angle of the smooth surface is $\theeq=145^\circ$, while for the
patterned plane $\theeq=100^\circ$ corresponding to a Cassie angle of 
$\theta_\mathrm{c}\approx145^\circ$. For $Bo > 0.3$ the drop on the smooth plane is very deformed from its equilibrium shape and a small further increase in $Bo$ leads to the drop detaching from the substrate. For the drop on the posts steady motion is only possible above a threshold Bond number due to pinning, but then the drop moves faster and with much less deformation. 
Thus it is possible to push the drop much harder and to achieve velocities larger by a factor $\sim 4$ before detachment occurs.

Increasing the bulk force $f$ leads to a drop that detaches from the surface.  
We have observed that, for superhydrophobic surfaces, there is a long 
transient regime in which the droplet accelerates before it finally detaches. 
This was not observed in the case of smooth planes.  
Similar accelerating  drops have been observed in experiments 
on superhydrophobic surfaces \cite{EXP2-ROLL} although the acceleration
measured in the simulations is much smaller ($0.2\% \times f$) than that 
in the experiments ($\sim f$).

\section{Discussion}\label{SecConclusions}

We have used a mesoscopic simulation approach to investigate the velocity profile of hydrophobic and
superhydrophobic drops subject to a constant body force.

For both hydrophobic and superhydrophobic surfaces
 we found that the velocity profile perpendicular to the substrate  
$v_x(z)$ comprises two regions. Close to the surface $v_x(z)$ is 
quadratic in $z$,
as in a Poiseuille flow.  This is the velocity profile found using 
lubrication theory \cite{Lub}.
Further from the substrate $v_x(z)$ is linear in $z$ describing 
a rolling motion, as predicted for drops of large $\theeq$ by scaling 
arguments \cite{MP-99,HJR-04}. The crossover
is clearly seen for equilibrium contact angles $\gtrapprox 120^\circ$, 
for smaller contact angles (e.g.\ $70^\circ$) the velocity profile is
quadratic for all $z$.

Fitting the linear portion of the curve as
\begin{eqnarray}
v_x &=& \vslip+ w\cdot z 
\end{eqnarray}
defines a mesoscopic slip velocity $\vslip$ which characterizes the quadratic
 region of the flow field. We find that $Ca_\mathrm{slip}$ is independent of the
viscosity and the surface tension 
but depends on the the shape 
of the drop, through the equilibrium contact angle,  and on
the mobility parameter $M$ which  controls the ability of 
the contact line to move across the surface.
 This is reminiscent of the 'inner' region, identified by Cox 
\cite{Cox}, which encapsulates the microscopic physics.

The rolling part of the flow profile was found to be independent of the mobility. Thus it might be identified as the 'outer' region of the
flow which is, as expected, independent of the microscopic details of the contact line motion. The angular velocity is independent of surface tension,
as expected, but increases with increasing viscosity.

A similar behaviour is seen for drops on superhydrophobic surfaces, but there is also a microscopic slip velocity, that is, a non-zero
velocity of the fluid adjacent to the surface. We stress that this occurs despite the no-slip boundary conditions
imposed in the simulations. It suggests that the contact line moves not by diffusion, but by tank-treading from one post to the next. Further evidence for this is that, for
the superhydrophobic surfaces, $\vslip$ is
 independent of the mobility.

Pinning on the posts leads to contact angle hysteresis and it takes a non-zero force to initiate drop motion.
Once the drop is moving it can be pushed more easily than on a flat surface and can reach higher velocities before flying off the substrate.
The drop slowly accelerates before leaving the surface.

Rolling dynamics have been observed in experiments.
In \cite{Richard-99} an almost pure rolling motion was measured 
for glycerol drops with a superhydrophobic coating ($\theeq=165^\circ$ with an
hysteresis $\Delta \theta\leq 10^\circ$) and diameters between
1.2 and 8 $mm$. The dominance of rolling was ascribed to the high 
viscosity of glycerol (950 times bigger than water) and the high equilibrium 
contact angle. Suzuki {\em et al.} \cite{EXP1-ROLL} reported a linear 
profile for the velocity  inside water drops with $\theeq\approx110^\circ$ 
but did not resolve any quadratic regime. However slip at the solid substrate 
was observed, and identified as $\vslip$, and three different
cases were reported in which $\vslip$ was negligible, comparable
and dominant with respect to the rolling part of the motion. In
\cite{EXP2-ROLL} predominantly sliding drops were observed in the 
superhydrophobic regime ($\theeq=150^\circ$), while for lower $\theeq$ 
rolling motions were also resolved.

A recent molecular dynamic simulation \cite{SM-08}  gave a Poiseuille profile (\ref{EqVQuad}) over all the height of the drop. In particular no rolling regime was observed. Possible explanations for the difference between these results and ours are that the equilibrium contact angle in the molecular dynamics simulations was $130^\circ$, corresponding to a region in our simulations where rolling was not well resolved, that molecular dynamics typically accesses much smaller length scales than the diffuse interface approach or that the intrinsic slip mechanism at the contact line was different in the two methods.

Clearly more work needs to be done to unify experiments, mesoscopic simulations and microscopic simulations to give a clear picture of how the way in which a drop moves across a surface depends on its geometry, the parameters describing the fluid and the way in which the drop interacts with the substrate. To work in this direction our next aim is to gain a better understanding of how the size of the drop and the degree of intrinsic slip at the surface affect the simulation results.
\\
\vspace{2.cm}
\\
{\bf ACKNOWLEDGMENTS}\\
We thank Rodrigo Ledesma and Matthew Blow for useful
discussions.

\end{document}